\documentclass[12pt]{article}
\usepackage{epsfig,amssymb,amsmath,psfrag}

\def\eqn#1{eq.~(\ref{#1})}

\def\eqns#1#2{eqs.~(\ref{#1}) and~(\ref{#2})}

\def\fig#1{fig.~{\ref{#1}}}

\def\Li{{\rm Li}}
\def\cN{{\mathcal N}}
\def\cO{{\mathcal O}}

\begin{document}
\thispagestyle{empty}

\begin{flushright}
HU-EP-11/17 \hskip1cm
CERN--PH--TH/2011-075\hskip1cm
SLAC--PUB--14434 \hskip1cm 
LAPTH-013/11
\end{flushright}

\begingroup\centering
{\Large\bfseries\mathversion{bold}
The one-loop six-dimensional hexagon integral\\
and its relation to MHV amplitudes in $\cN=4$ SYM \par}%
\vspace{7mm}

\begingroup\scshape\large
Lance~J.~Dixon,
\endgroup
\vspace{3mm}

\begingroup
\textit{PH-TH Division, CERN, Geneva, Switzerland 
 }\\
 \vspace{3mm}
 and\\
 \vspace{3mm}
\textit{SLAC National Accelerator Laboratory, \\
Stanford University, Stanford, CA 94309, USA 
   }\\ 
\par
\texttt{lance@slac.stanford.edu\phantom{\ldots}}
\endgroup

\vspace{1.0cm}

\begingroup\scshape\large
James~M.~Drummond,
\endgroup
\vspace{3mm}

\begingroup
\textit{PH-TH Division, CERN, Geneva, Switzerland 
 }\\
\vspace{3mm}
 and\\
 \vspace{3mm}
\textit{LAPTH, Universit\'e de Savoie, CNRS \\
B.P. 110, F-74941 Annecy-le-Vieux Cedex, France
   }\\ 
\par
\texttt{drummond@lapp.in2p3.fr\phantom{\ldots}}
\endgroup

\vspace{1.0cm}

\begingroup\scshape\large
Johannes M.~Henn
\endgroup
\vspace{3mm}

\begingroup
\textit{Institut f\"ur Physik, Humboldt-Universit\"at zu Berlin, \\
Newtonstra{\ss}e 15, D-12489 Berlin, Germany}\par
\texttt{henn@physik.hu-berlin.de\phantom{\ldots}}
\endgroup

\vspace{0.5cm}

\textbf{Abstract}\vspace{5mm}\par
\begin{minipage}{14.7cm}
We provide an analytic formula for the
(rescaled) one-loop scalar hexagon integral $\tilde\Phi_6$ with all
external legs massless, in terms of classical polylogarithms.  We show
that this integral is closely connected to two integrals appearing in
one- and two-loop amplitudes in planar $\cN=4$ super-Yang-Mills
theory, $\Omega^{(1)}$ and $\Omega^{(2)}$.  The derivative of
$\Omega^{(2)}$ with respect to one of the conformal invariants
yields $\tilde\Phi_6$, while another first-order differential operator
applied to $\tilde\Phi_6$ yields $\Omega^{(1)}$.  We also introduce
some kinematic variables that rationalize the arguments of the
polylogarithms, making it easy to verify the latter differential equation.
We also give a further example of a six-dimensional integral relevant
for amplitudes in $\cN=4$ super-Yang-Mills.
\end{minipage}\par
\endgroup

\newpage

%\tableofcontents

\section{Introduction and outline}

Recent years have seen dramatic progress in the understanding of
multi-loop and multi-leg scattering amplitudes in $\cN=4$ super
Yang-Mills theory (SYM), especially in the planar limit.  The planar
amplitudes have a hidden dual conformal 
symmetry~\cite{Drummond:2006rz,Bern:2006ew,Drummond:2007cf} that leads to
powerful constraints.  There is also a surprising correspondence
between scattering amplitudes and Wilson loops
\cite{Alday:2007hr,Drummond:2007aua,Brandhuber:2007yx}; see
refs.~\cite{Mason:2010yk,CaronHuot:2010ek,Belitsky:2011zm,Gaiotto:2011dt}
for recent developments.  
A dual conformal Ward identity~\cite{Drummond:2007au}, derived for
Wilson loops, can be used to fix the functional form of multi-loop
scattering amplitudes, up to {\it a priori} undetermined functions
of dual conformal cross-ratios.  For example, the functional form of
the four- and five-point amplitudes is uniquely fixed to all orders
in the coupling
constant, in agreement with explicit computations in field
theory~\cite{Bern:2006ew,Anastasiou:2003kj,Bern:2005iz,Cachazo:2006tj,%
Bern:2006vw,Bern:2008ap,Cachazo:2008hp,Spradlin:2008uu,Henn:2010ir}
and string theory~\cite{Alday:2007hr}.  For maximally-helicity-violating
(MHV) amplitudes, the difference between the (logarithms of the)
particular solution to the Ward identity (the BDS ansatz~\cite{Bern:2005iz})
and the amplitude is called the remainder
function~\cite{Bern:2008ap,Drummond:2008aq}.  For six external particles,
this remainder function can depend only on three dual conformal cross ratios
$u_1,u_2$ and $u_3$.

Another important consequence of dual conformal symmetry is a powerful
restriction on the planar loop integrand, which had been observed in
dimensional regularization~\cite{Drummond:2006rz,Bern:2006ew,Bern:2007ct},
and can be made rigorous on the Coulomb branch of $\cN=4$ SYM
\cite{Henn:2010kb,Dennen:2010dh,CaronHuot:2010rj,Henn:2011xk}.

The six-point remainder function at two loops is known
analytically~\cite{DelDuca:2009au,DelDuca:2010zg,Goncharov:2010jf},
thanks to the correspondence between scattering amplitudes and Wilson loops.
On the amplitude side, so far results are available
numerically~\cite{Bern:2008ap} and
analytically in certain kinematical
limits~\cite{Bartels:2008ce,Bartels:2008sc,Drummond:2010mb}.
Recently, iterative differential equations were used to
directly evaluate integrals that contribute to the scattering
amplitudes~\cite{Drummond:2010cz}.

The motivation of the present paper is to show how to derive
analytical results for loop integrals relevant for multi-leg
scattering amplitudes, using differential equations.
We concentrate on the six-point case, but our
method is also applicable to more external legs.

The ``even'' part of the planar
six-particle MHV scattering amplitude at two
loops was first given in ref.~\cite{Bern:2008ap} in terms of
fifteen separate integrals with simple dual conformal properties.
It can be represented alternatively~\cite{Drummond:2010mb,ArkaniHamed:2010kv}
in terms of six dual conformal two-loop integrals,
five of which are infrared divergent and one of which is finite.
The finite integral, denoted by
$\Omega^{(2)}$, depends on the three dual conformal cross-ratios
$u_1 , u_2 , u_3 $.  It is reasonable to believe that it contains an
essential part of the two-loop six-point remainder function. 
In ref.~\cite{Drummond:2010cz} it was found that $\Omega^{(2)}$ satisfies
several simple second-order differential equations, one of which
relates it to an analogous one-loop integral, called $\Omega^{(1)}$.

\begin{figure}
\centerline{ 
{\epsfxsize12cm  \epsfbox{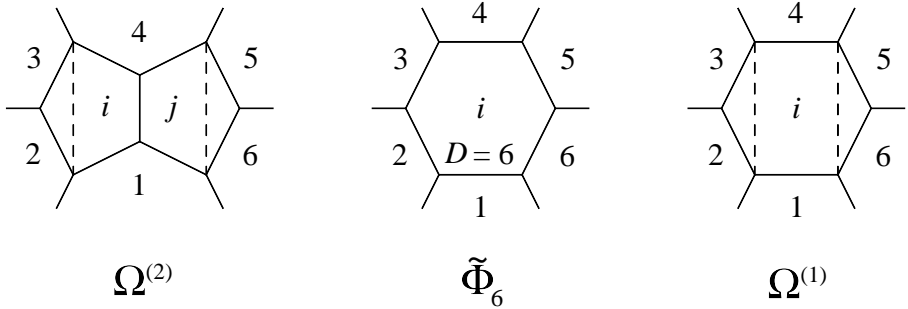}}
}
\caption{\small
Three dual conformal integrals which are related to each other by the
action of first-order differential operators, as discussed in the text.
The labels $i,j,1,2,\ldots,6$ are indices $k$ for dual (or region)
coordinates $x_k$.  Solid lines indicate propagators; dashed lines
indicate numerator factors of $x_{ai}^2$ or $x_{bi}^2$, as explained
in the text.  The central integral $\tilde{\Phi}_6$ has no such numerator
factors, but is evaluated in dimension $D=6$ instead of $D=4$.  The standard
hexagon integral $H$ is rescaled to obtain a dual conformal invariant
integral $\Phi_6$, which is rescaled once again to obtain the pure
degree 3 function $\tilde\Phi_6$.}
\label{fig-threeintegrals}
\end{figure}

In this paper we observe that the one-loop scalar hexagon
integral in six space-time dimensions is related to the
aforementioned four-dimensional integrals via first-order differential
equations. The relations that we find are (schematically)
\begin{eqnarray}\label{intro-omega-phi-omega}
\Omega^{(2)}(u_1, u_2 , u_3 )
\longrightarrow \tilde\Phi_{6} (u_1, u_2 , u_3 )
\longrightarrow \Omega^{(1)} (u_1, u_2 , u_3 )\,,
\end{eqnarray}
where the arrows denote certain first-order differential operators in
the $u_i$.  (See \fig{fig-threeintegrals}.)
Here $\tilde\Phi_{6}$ stands for the six-dimensional scalar
hexagon integral, after two simple rescalings. The first (to $\Phi_6$)
makes it invariant under dual conformal transformations.  The second
removes an algebraic prefactor.  It is natural to consider
$\tilde\Phi_{6}$ as an intermediate step between $\Omega^{(1)}$ and
$\Omega^{(2)}$.  Thanks to the high degree of symmetry of the
hexagon integral,
the first-order differential equation relating $\tilde\Phi_{6}$ 
(or $\Phi_6$) and
$\Omega^{(1)}$ in fact leads to a system of three inequivalent
equations.  Together with a simple boundary condition, the latter
completely determines the three-variable function
$\Phi_{6}( u_1, u_2 , u_3 )$.

%%%%%%%%%%%%%%%%%%%%%%%%%%%%%%%%%%%%%%%%%
%%%%%%%%%%%%%%%%%%%%%%%%%%%%%%%%%%%%%%%%%

From a practical viewpoint, the intermediate step between
$\Omega^{(2)}$ and $\Omega^{(1)}$ in \eqn{intro-omega-phi-omega} is
very useful. It is also very natural, since the $\Omega^{(i)}$
functions are expected to be given by linear combinations of functions
defined through iterated (poly)logarithmic integrals, such as
$\log^n$, $\Li_{n}$, and generalizations thereof.  If we associate a
``degree of transcendentality'' with the number of iterated integrals,
then $\Omega^{(1)}$, $\tilde\Phi_{6}$ and $\Omega^{(2)}$ are pure
functions of degree $2, 3$ and $4$, respectively.  In some sense,
$\tilde\Phi_{6}$ represents a ``one-and-a-half'' loop function.

We find that the solution for $\Phi_{6}$ is given by a simple formula
in terms of degree three functions, \eqn{phitfullsolution} below.
It is remarkably similar in structure to the two-loop remainder function.

The six-dimensional hexagon integral $\Phi_{6}$ also is of inherent interest
for a number of reasons. In dimensional regularization with
$4-2\epsilon$ dimensions, it appears in the $\cO(\epsilon)$ part of
the one-loop six-particle MHV amplitude~\cite{Bern:1996ja}.
It is generated because a term in the numerator of the one-loop
integrand contains a factor
of $\ell_{[-2\epsilon]}^2 \equiv \mu^2$, where $\ell_{[-2\epsilon]}$
denotes the components of the loop momenta that lie outside of four
dimensions.  The integral of such a term yields $-\epsilon$ times
the scalar integral in six dimensions.  Moreover,
in order to determine the remainder function at higher loops, one has
to take the logarithm of the amplitude, in which case
$\cO(\epsilon^{i})$ terms at lower loops get multiplied by pole terms
$\epsilon^{-j}$ (with $j\le2L$, where $L$ is the loop order.)  The
$\cO(\epsilon)$ terms must
be kept in order to obtain a consistent result at $\cO(1)$. As an
example, when computing the two-loop remainder function in this way,
$\Phi_{6}$ participates in a cancellation involving certain two-loop
``hexabox'' integrals~\cite{Bern:2008ap}, where again there is a 
factor of $\mu^2$ in the numerator for the hexagon loop.
This link between higher-order terms in
the $\epsilon$ expansion and higher-loop integrals also motivates
the idea that $\Phi_{6}$ should already know about some of the structure
of the two-loop answer, and our result supports this expectation.

Another motivation for considering six-dimensional integrals in
general is the known connection between scalar integrals in $(D+2)$
dimensions and tensor integrals in $D$ dimensions (see
e.g.~ref.~\cite{Bern:1993kr}.)  In particular, many of the finite tensor
integrals introduced in ref.~\cite{ArkaniHamed:2010kv} can be viewed as
higher-dimensional scalar integrals, or are related to them via
differential equations.  This relation does not depend on dual
conformal symmetry.  As an example, we will show a six-dimensional
integral, and equivalently, a four-dimensional tensor integral, that
computes the finite part of the two-mass-easy box integral.

The hexagon integral $\Phi_{6}$ is a function of three dual
conformally invariant cross-ratios $u_{1},u_{2},u_{3}$.  Like the
two-loop remainder function, it is conveniently expressed in terms of
a set of redundant variables $x_{i \pm} = u_{i} x_{\pm}$, where
\begin{equation}
\label{xpmDelta}
x_{\pm} = \frac{-1+u_{1}+u_{2}+u_{3} \pm \sqrt{\Delta}}{2 u_1 u_2 u_3 } \,, 
\qquad \Delta = (-1+u_1 + u_2 + u_3 )^2 - 4 u_1 u_2 u_3 \,.
\end{equation}
Later we will give a change of variables from $u_{i}$ to a set of
variables $v_{0}, v_{\pm}$.  Although these variables do not manifest
the cyclic symmetry, they have the feature that the arguments of the
polylogarithms in the result for the hexagon integral,
and also all terms in the differential equations,
are rational functions of $v_{0}, v_{\pm}$, with no square roots.
This is very convenient for verifying the differential equations. Analogous
transformations may be also useful when considering other six-point integrals.

Recently, the notion of symbols was advocated as a tool to think about
iterated integrals appearing in $\cN=4$ SYM
\cite{Goncharov:2010jf}. We compute the symbol of the hexagon integral
and find that it is given by a very simple expression. Its simplicity
follows from the differential equations that $\Phi_{6}$ satisfies.

This paper is organized as follows. We begin by defining the hexagon
integral $\Phi_{6}$ and discussing its symmetry properties in section
\ref{sect-preliminaries}. We also explain how dual conformal symmetry
helps in obtaining a simple Feynman parametrization, which is a
general feature. Another example is given in Appendix
\ref{appendix-wl}. We then point out the relation between $\Phi_{6}$
and integrals appearing in the two-loop six-point
MHV amplitude in $\cN=4$ SYM, in the representation of
ref.~\cite{ArkaniHamed:2010kv}.
This relation takes the form of first-order differential
equations. We present the analytic solution to these equations in
section \ref{sect-analytic-solution}.  In section \ref{sect-variables}
we introduce a convenient set of variables that renders the arguments
of the functions appearing in $\Phi_{6}$ rational, and directly verify
the differential equations. In section \ref{sect-symbol} we discuss
the symbol of $\Phi_{6}$. We conclude and give an outlook in section
\ref{sect-conclusion}.
 
%%%%%%%%%%%%%%%%%%%%%%%%%%%%%%%%%

\section{Six-dimensional hexagon integral}

\subsection{Preliminaries}
\label{sect-preliminaries}

We consider the on-shell six-dimensional scalar hexagon integral
$H$ in $D=6$ dimensions, with external momenta $p^{\mu}_{j}$ satisfying
momentum conservation, $\sum_{j=1}^6 p^{\mu}_{j} = 0$, and masslessness,
$p_j^2 = 0$ for $j=1,2,\ldots,6$. In terms of dual (or region) coordinates
$p^{\mu}_{j} = x^{\mu}_{j} - x^{\mu}_{j+1}$, it is defined by
\begin{eqnarray}\label{def-H}
H  = \int \frac{d^{6}x_{i}}{i \pi^3} \frac{1}{\prod_{j=1}^{6} x_{ij}^2}\,,
\end{eqnarray}
where $x^{\mu}_{ij} = x^{\mu}_{i} - x^{\mu}_{j}$, and $x_i^\mu$ is
the dual coordinate corresponding to the loop momentum
(see~\fig{fig-threeintegrals} for the labeling).
The integral is both ultraviolet (UV) and infrared (IR) finite.
As a scalar integral, $H$ is a function of the external Lorentz invariants 
$x_{j,j+2}^2 = s_{j,j+1}$ and $x_{j,j+3}^2 = s_{j,j+1,j+2}$.
Here $s_{j,j+1} = (p_{j}+p_{j+1})^2$ and
$s_{j,j+1,j+2} = (p_{j} + p_{j+1}+p_{j+2})^2$, and external indices
are defined modulo 6.  We work in signature $({-}{+}{+}{+})$, so that
the Euclidean region has all $s_{j,j+1}$ and $s_{j,j+1,j+2}$ positive.
The on-shell conditions, $p_{j}^2=0$,
are expressed in dual coordinates as $x_{j,j+1}^2=0$. 
Momentum conservation translates to $x^\mu_{j+6} \equiv x^{\mu}_{j}$
in the dual space.

Covariance of $H$ under dual conformal
symmetry~\cite{Broadhurst:1993ib,Drummond:2006rz},
in particular under the inversion of all dual coordinates,
$x^{\mu} \to x^{\mu}/x^2$, allows us to write
\begin{align}\label{phi-crossratios}
s_{123} s_{234} s_{345} H(s_{i,i+1}, s_{i,i+1,i+2})
\ \equiv\ \Phi_{6}( u_1 , u_2 , u_3 )\,,
\end{align}
where the cross-ratios
\begin{equation}
\label{defu}
u_{1} = \frac{x_{13}^2 x_{46}^2}{x_{14}^2 x_{36}^2}
 = \frac{s_{12} s_{45}}{s_{123} s_{345}}\,, \qquad
u_{2} = \frac{x_{24}^2 x_{51}^2}{x_{25}^2 x_{41}^2}
 = \frac{s_{23} s_{56}}{s_{234} s_{123}}\,, \qquad
u_{3} = \frac{x_{35}^2 x_{62}^2}{x_{36}^2 x_{52}^2}
 = \frac{s_{34} s_{61}}{s_{345} s_{234}}\,,
\end{equation}
are invariant under dual conformal transformations.

We observe that $\Phi_{6}$ has both cyclic and reflection symmetries.
This leads to a full permutation symmetry in the
$\{u_{1},u_{2},u_{3}\}$, i.e.
\begin{eqnarray}\label{phi-permutation-symmetry}
\Phi_{6}(u_{1},u_{2},u_{3}) = \Phi_{6}(u_{3},u_{1},u_{2})
 = \Phi_{6}(u_{2},u_{3},u_{1})\,,\qquad
\Phi_{6}(u_{1},u_{2},u_{3}) =\Phi_{6}(u_{2},u_{1},u_{3}) \,.
\end{eqnarray}
We will compute $\Phi_{6}$ in the Euclidean region, i.e.~where
$s_{j,j+1}>0$,\ \ $s_{j,j+1,j+2}>0$.  Although we will eventually compute
$\Phi_{6}$ from differential equations, it is useful to have a simple
parametric representation for $\Phi_{6}$, for example for numerical checks.
Here we give an instructive example that highlights technical
simplifications brought about by dual conformal symmetry that may be of
more general interest.

Introducing Feynman parameters in the standard
way~\cite{smirnov2006feynman}, we have
\begin{eqnarray}\label{phi-feynman}
\Phi_{6}(u_{1}, u_{2} , u_{3} )
= 2 \,x_{14}^2 x_{25}^2 x_{36}^2
\int_{0}^{\infty} \prod_{i=1}^6 d\alpha_{i} \, 
\frac{\delta(\sum_{i=1}^{6} c_{i} \alpha_{i} -1) }
{\left[ \sum_{i<j} x_{ij}^2 \alpha_{i} \alpha_{j} \right]^3}\,.
\end{eqnarray}
Note that we can choose the $c_{i}$ arbitrarily, as long as at least
one of them is different from zero~\cite{smirnov2006feynman}.

We have already seen that dual conformal symmetry leads to the
simplified variable dependence~(\ref{phi-crossratios}). Moreover, dual
conformal symmetry often leads to further simplifications in the
evaluation of loop integrals. For example, it is well
known~\cite{Broadhurst:1993ib} that in the off-shell case, a combination of
a translation and an inversion in the dual space of the $x_{i}$ can be
used to send one of the dual points to infinity, thereby reducing the
number of propagators by one.  In this way, Broadhurst demonstrated
the equivalence of an infinite class of off-shell three- and four-point
ladder integrals.

In the present case, we cannot immediately use the same idea, due to
the light-like constraints $p_{j}^2 = x_{j,j+1}^2 = 0$, which
would make the above-mentioned inversion singular. However, we can
nevertheless exploit technical simplifications that dual conformal
symmetry entails.

For a generic one-loop integral, a factor of 
$(\sum_{i} \alpha_{i})^{a-D}$, where $a$ is the number of propagators,
would be present under the integral sign on the right-hand side
of~\eqn{phi-feynman}.
Here, this factor is absent since $a=D=6$, which is precisely the
condition for dual conformal symmetry.  In this case it is often
convenient to choose one or more $c_{i}=0$, because the resulting
integrals from $0$ to $\infty$ in \eqn{phi-feynman} are easy to
carry out.  We will set $c_{6}=c_{1}=c_{2}=0$, $c_{3}=c_{4}=c_{5}=1$.
We will also use the redundancy in \eqn{defu} to set
$x_{46}^2=u_{1}, x_{24}^2 = u_{2}, x^2_{26} = u_{3}$.  (All other
$x_{jk}^2$ appearing in \eqn{defu} are set to 1.)

Performing the $\alpha_{6},\alpha_{1},\alpha_{2}$ integrations,
we readily obtain
\begin{equation}\label{phi-feynman2}
\Phi_{6}(u_{1}, u_{2} , u_{3} )
= \int_{0}^{1} d\alpha_{3,4,5}  \log\left(\frac{a d}{bc} \right)
\frac{\delta(\sum_{i=3}^{5} \alpha_{i} -1)}{ad-bc} \,,
\end{equation} 
where $ad = \alpha_3 \alpha_5 u_{3}$, $b= \alpha_4 u_{2} + \alpha_5 $
and $c=\alpha_4 u_{1} + \alpha_3 $.  In this form, it is easy to see that
the answer will be built from degree three functions.

%%%%%%%%%%%%%%%%%
\subsection{Relation to integrals appearing in the six-point MHV amplitude}

As was mentioned in the introduction, $\Phi_{6}$ appears in the
$O(\epsilon)$ part of the one-loop six-particle MHV amplitude in
dimensional regularization~\cite{Bern:1996ja}.  Moreover, when
computing the logarithm of that amplitude to two loops, $\Phi_{6}$
participates in a cancellation involving certain two-loop hexabox
integrals~\cite{Bern:2008ap}. It is therefore not unreasonable to
think that $\Phi_{6}$ already contains some of the structure of the two-loop
result.

In fact, one can find a very direct relation between integrals
relevant for MHV scattering amplitudes and $\Phi_{6}$. In
refs.~\cite{ArkaniHamed:2010kv,ArkaniHamed:2010gh}, dual conformal integrals
with a tensor structure in the numerator were introduced for the
description of scattering amplitudes in $\cN=4$ SYM.  One of them is
given by
\begin{eqnarray}\label{def-omega1}
\Omega^{(1)}(u_1 , u_2 ,u_3 )
= - \frac{x_{35}^2 x_{26}^2 x_{14}^2}{x_{ab}^2} 
 \int \frac{d^{4}x_{i}}{i \pi^2}
\frac{x_{ai}^2 x_{bi}^2}{\prod_{j=1}^{6} x_{ji}^2}\,,
\end{eqnarray} 
where $x_{a}^{\mu}$ is a solution to the four-cut condition
$x_{1a}^2=x_{2a}^2=x_{3a}^2=x_{4a}^2=0$, and $x_{b}^{\mu}$ is obtained
from $x_{a}^{\mu}$ by a rotation by $3$ units (see \fig{fig-threeintegrals}).
The two choices for $x_a$ are related by parity. For the finite integrals we consider here the result is independent of the choice. The numerator factor
$x_{ai}^2 x_{bi}^2$ is crucial in order to make the integral IR
finite~\cite{ArkaniHamed:2010kv,ArkaniHamed:2010gh}.  The definition of the
numerator and the normalization in~\eqn{def-omega1} are easy to
write out explicitly in twistor-space notation.  We refer the
interested reader to refs.~\cite{ArkaniHamed:2010kv,Drummond:2010cz} for
further details.

One might think that $\Omega^{(1)}$ would be a rather complicated hexagon
integral.  However, dual conformal symmetry and the specific choice of
the numerator in~\eqn{def-omega1} allow it to be given
by a remarkably simple formula,
\begin{eqnarray}
\label{Om1}
\Omega^{(1)}(u_{1},u_{2},u_{3}) = \log u_{1} \log u_{2} 
+ {\rm Li}_{2}(1-u_{1}) + {\rm Li}_{2}(1-u_{2})
+ {\rm Li}_{2}(1-u_{3}) - 2 \zeta_{2}\,.
\end{eqnarray}

The integral $\Omega^{(1)}$ also plays an important role as the source
term for a second-order differential equation for $\Omega^{(2)}$, an integral
appearing in the two-loop six-particle MHV amplitude~\cite{Drummond:2010cz}.
The latter integral is defined by 
\begin{eqnarray}\label{def-omega2}
\Omega^{(2)}(u_1 , u_2 ,u_3 ) =
- \frac{x_{35}^2 x_{26}^2 (x_{14}^2)^2}{x_{ab}^2} 
\int \frac{d^{4}x_{i}}{i \pi^2} \int \frac{d^{4}x_{j}}{i \pi^2} 
\frac{x_{ai}^2 x_{bj}^2}{x_{1i}^2 x_{2i}^2 x_{3i}^2 x_{4i}^2
x_{ij}^2 x_{4j}^2 x_{5j}^2 x_{6j}^2 x_{1j}^2}\,,
\end{eqnarray} 
where the definition of $x_{a}^{\mu}$ and $x_{b}^{\mu}$ is the same
as for $\Omega^{(1)}$ in \eqn{def-omega1}.  This integrals is also
depicted in~\fig{fig-threeintegrals}. 

The differential equation obeyed by $\Omega^{(2)}$ is~\cite{Drummond:2010cz}
\begin{eqnarray}\label{diffeq-phi2-phi1}
u_{3} \partial_{u_{3}} \tilde{D}^{(1)}  \Omega^{(2)} = \Omega^{(1)}\,,
\end{eqnarray}
where $\tilde{D}^{(1)}$ is the first-order differential operator
\begin{eqnarray}
\label{def-tD1}
\tilde{D}^{(1)} = - u_{1} (1-u_{1}) \partial_{u_{1}}
                  - u_{2} (1-u_{2}) \partial_{u_{2}}
                  + (1-u_{1}-u_{2}) (1-u_{3})  \partial_{u_{3}}  \,.
\end{eqnarray}
Given the factorized structure of the second-order differential operator
in \eqn{diffeq-phi2-phi1}, it is natural to search for an object which
sits ``between'' $\Omega^{(2)}$ and $\Omega^{(1)}$.  The $D=6$ scalar
hexagon integral, with transcendentality degree 3, is
a particularly good candidate for such an object.

Inspecting the Feynman parametrization\footnote{J.~M.~Henn thanks 
N.~Arkani-Hamed and J.~Bourjaily for collaboration on Feynman
parametrizations of twistor integrals.}
of $\Omega^{(1)}$, it is easy to see that it
is related to $\Phi_{6}$ in the following way,
\begin{equation}\label{link-phi-omega1}
D^{(1)} \Phi_{6} = \Omega^{(1)}  \,,
\end{equation}
where $D^{(1)}$ is the first-order differential operator
\begin{equation}\label{def-D1}
D^{(1)} =   \frac{u_{3}}{u_{1} u_{2}} \left[
   u_{1} (1-u_{1}) \partial_{u_{1}} 
 + u_{2} (1-u_{2}) \partial_{u_{2}}
 - (1-u_{1}-u_{2}) (1-u_{3}) \partial_{u_{3}}
 - 1 \right] u_{1} u_{2} \,.
\end{equation}
This relation is not particularly surprising, since it is well known that
tensor integrals in $D$ dimensions are often related to scalar
integrals in $(D+2)$ dimensions~\cite{Bern:1993kr}. We give a further
example in Appendix \ref{4d-6d-connection}. 
Relation~(\ref{link-phi-omega1}) is easy to understand: when acting on the
scalar integrand of $\Phi_{6}$ in Feynman parameter form, see
\eqn{def-H}, the differential operator~(\ref{def-D1}) creates terms
that are equivalent to those coming from the numerator of
$\Omega^{(1)}$. Further, the increase in the power of the denominator
due to the differentiation can be absorbed by a shift in the dimension
from $6$ to $4$.

Let us comment further on the remarkable
link between $\Phi_{6}$ and $\Omega^{(2)}$.  
We can commute the two first-order
operators in \eqn{diffeq-phi2-phi1}.  Using
\begin{eqnarray}
\left[u_{3} \partial_{u_3},  \tilde{D}^{(1)} \right]
 &=& - (1 - u_{1} - u_{2}) \partial_{u_3} \,, \\
 D^{(1)} &=& -\tilde{D}^{(1)} u_3 + ( 1 - u_1 - u_2 ) \,,
\end{eqnarray}
we have
\begin{eqnarray}
\label{new2ndorderdiffeq}
{D}^{(1)} \partial_{u_{3}}  \Omega^{(2)}\ =\ - \Omega^{(1)} \,.
\end{eqnarray}
Comparing \eqn{new2ndorderdiffeq} with \eqn{link-phi-omega1},
we find that
\begin{equation}
\partial_{u_3} \Omega^{(2)}\ =\ -\Phi_6 + K \,,
\end{equation}
where $K$ satisfies $D^{(1)} K = 0$.  In fact we find numerically
that $K=0$.
Thus $\Phi_{6}$ can be considered as an intermediate step
between $\Omega^{(1)}$ and $\Omega^{(2)}$.  Only one more integration
of $\Phi_{6}$ is required to obtain $\Omega^{(2)}$.
Consistent with these differential equations, the degree of
transcendentality increases from $\Omega^{(1)}$ to $\Phi_{6}$ to
$\Omega^{(2)}$ in steps of one. Considering its links to the six-particle MHV amplitudes in $\mathcal{N}=4$ super Yang-Mills, it is of interest to understand better the function $\Phi_6$.

Let us proceed to evaluate the hexagon integral.
The idea is to use \eqn{link-phi-omega1} in order to determine
$\Phi_{6}$. We will first put the equation into a more useful form.
The zeroth-order piece in \eqn{def-D1} suggests that $\Phi_{6}$
has some algebraic prefactor.  Indeed, let us define
\begin{equation}
\tilde{\Phi}_{6}\ :=\ \sqrt{\Delta} \, \Phi_{6} \,,
\end{equation}
where $\Delta = (u_{1}+u_{2}+u_{3}-1)^2 -4 u_{1} u_{2} u_{3}$.
Then, thanks to 
$D^{(1)} ({1}/{\sqrt{\Delta}}) = 0 $,
it is straightforward to commute the first-order part of $D^{(1)}$
around $u_{1} u_{2} /\sqrt{\Delta}$,
and one obtains 
\begin{eqnarray} \label{diffeqphitilde}
- \frac{u_{3}}{\sqrt{\Delta}} \tilde{D}^{(1)} \tilde{\Phi}_{6}
 = \Omega^{(1)}\,,
\end{eqnarray}
where the operator $\tilde{D}^{(1)}$ given in \eqn{def-tD1}
no longer contains zeroth-order terms.
Due to the permutation symmetry~(\ref{phi-permutation-symmetry})
in the arguments of $\Phi_{6}$, \eqn{diffeqphitilde}
leads to two further non-trivial first-order differential equations. 
This set of differential equations determines $\Phi_{6}$ up
to one integration constant. The latter can be fixed by the requirement
that $\Phi_{6}$ should be non-singular at $\Delta=0$, which
implies the vanishing of $\tilde{\Phi}_{6}$ on that locus.
 
Diagonalizing the set of differential equations generated
by~\eqn{diffeqphitilde}, we have
\begin{eqnarray}\label{diffeqsinglevar1}
\partial_{u_{1}} \tilde{\Phi}_{6}(u_{1},u_{2},u_{3}) &=&  \null
- \frac{ 1-u_{1}+u_{2}-u_{3}}{(1-u_{1}) \sqrt{ \Delta }}
\Omega^{(1)}(u_{1},u_{2},u_{3}) \\
&& \null - \frac{ 1-u_{1}-u_{2}-u_{3}}{u_{1} \sqrt{ \Delta } }
\Omega^{(1)}(u_{2},u_{3},u_{1})
 - \frac{ 1-u_{1}-u_{2}+u_{3}}{{(1-u_{1}) \sqrt{ \Delta }}}
\Omega^{(1)}(u_{3},u_{1},u_{2})  \,, \nonumber
\end{eqnarray}
plus the two cyclically related equations.  In the next subsection,
we will present the full solution for $\Phi_{6}(u_{1},u_{2},u_{3})$.

%%%%%%%%%%%%%%%%%%%%%%%%%%%%%%%%%%%%%%%%%

\subsection{Result for ${\Phi}_{6}(u_{1},u_{2},u_{3})$}
\label{sect-analytic-solution}
Here we present the solution to the differential
equations~(\ref{diffeqphitilde}), or equivalently~(\ref{diffeqsinglevar1}).
We first define the variables
\begin{equation}
\label{xipm}
x_{i \pm} = u_{i} x_{\pm} \,,
\end{equation}
where $x_{\pm}$ and $\Delta$ are given in \eqn{xpmDelta}.
The appearance of the $x_{i \pm}$ should not come as a surprise,
since they played a prominent role in the two-loop remainder
function~\cite{Goncharov:2010jf}, and we have already argued that
$\Phi_{6}$ should capture some of its structure. 

Further, we define
\begin{eqnarray}
\label{funnyfunctions}
L_{3}(x_{+},x_{-})
&=& \sum_{m=0}^{2} \frac{(-1)^m}{(2m)!!}  \log^m (x_{+} x_{-})
\left[ \ell_{3-m}(x_{+}) - \ell_{3-m}(x_{-}) \right]\,,\\
\ell_{m}(x) &=& \frac{1}{2} ( \Li_{m}(x)-(-1)^{m} \Li_{m}(1/x) ) \,,
\end{eqnarray}
which is very similar to the function $L_{4}$ defined in
ref.~\cite{Goncharov:2010jf}.  As in ref.~\cite{Goncharov:2010jf},
the branch cuts of $\Li_{n}(x_{+})$ and $\Li_{n}(1/x_{-})$ are taken
to lie below the real axis,
i.e.~$\Li_{n}(x_{+}) := \Li_{n}(x_{+} + i \epsilon)$, etc.,
and the branch cuts of $\Li_{n}(x_{-})$ and $\Li_{n}(1/x_{+})$
are taken to lie above the real axis.\footnote{%
We are grateful to M.~Spradlin and C.~Vergu for discussions and
correspondence on the branch cut structure of $L_{4}$ in
ref.~\cite{Goncharov:2010jf}.}

We found the following formula for ${\Phi}_{6}$,
\begin{eqnarray}\label{phitfullsolution}
{\Phi}_{6}(u_{1},u_{2},u_{3})
\ =\ \frac{ {\tilde\Phi}_{6}(u_{1},u_{2},u_{3}) }{\sqrt{\Delta}}
\ =\ \frac{1}{\sqrt{\Delta}} \left[
- 2 \sum_{i=1}^{3} L_{3}( x_{i +},x_{i -})  
+ 2 \zeta_{2} J 
+ \frac{1}{3}  J^3 \right] 
\,,
\end{eqnarray}
where
\begin{eqnarray}
\label{def-J}
J\ =\ \sum_{i=1}^{3}  \left[ \ell_{1}(x_{i+}) - \ell_{1}(x_{i-}) \right]\,.
\end{eqnarray}
Although individual terms in~\eqn{phitfullsolution} can be complex,
their sum is always real in the Euclidean region $u_{i}>0$.

In the next section, we prove directly that \eqn{phitfullsolution}
satisfies the differential equations~(\ref{diffeqphitilde}).
In section \ref{sect-symbol}, we will see another way to
justify~\eqn{phitfullsolution} based on the differential
equations for its symbol.

%%%%%%%%%%%%%%%%
\subsection{Direct verification of the differential equations}
\label{sect-variables}

We found the following change of variables to be convenient,
\begin{eqnarray}\label{changevars}
u_1 = \frac{ v_{0} - v_{+} v_{-}  }{1+v_{0}-v_{+}-v_{-} } \,, \quad
u_2 =  \frac{ v_{0} - v_{+} v_{-}  }{(1+v_{0}-v_{+}-v_{-} ) v_{0}} \,, 
\quad
u_{3} = \frac{v_{+} v_{-}}{v_{0}} \,. 
\end{eqnarray}
This definition is symmetric in $v_{+}$ and $v_{-}$.  Choosing
$v_{+}>v_{-}$ without loss of generality, the inverse transformation
is given by
\begin{eqnarray}
v_{+} = u_{1} u_{3} x_{+} \,, \quad
v_{-} = u_{1} u_{3} x_{-} \,,\quad
v_{0} = \frac{u_1 }{u_2 } \,.
\end{eqnarray}
We also have the following useful expressions for the $x_{i \pm}$,
\begin{equation}
\label{xipmfromvi}
x_{1 \pm} = \frac{v_0}{v_\mp} \,, \quad
x_{2 \pm} = \frac{1}{v_\mp} \,, \quad
x_{3 \pm} = \frac{v_\pm (1 + v_0 - v_{+} - v_{-})}{v_0 - v_{+}v_{-}} \,.
\end{equation}
In terms of the variables $v_{0,+,-}$, $\Delta$ is a perfect square,
\begin{equation}
\Delta = \frac{(v_{+} - v_{-})^2 (v_{0} - v_{+} v_{-})^2}
    {(1+v_{0}-v_{+}-v_{-})^2 v_{0}^2} \,.
\end{equation}
In the Euclidean region $u_{i}>0$ that we are considering,
we can take the square root $\sqrt{\Delta}$ without
sign ambiguities, see \eqn{changevars}.  

In the remainder of this section, we will assume $\Delta>0$
for simplicity, so that the $v_{\pm}$ are real.
Note that the factor $J$
defined in \eqn{def-J} becomes simply
\begin{equation}
\label{Jinvi}
J\ =\ -\frac{1}{2} \, \log \frac{v_+}{v_-} \,.
\end{equation}

The differential equations~(\ref{diffeqsinglevar1}) are easily expressed
in the new variables, using Jacobian factors such as 
\begin{eqnarray}
\frac{\partial u_{1}}{\partial v_{+}}
= \frac{(v_{0} - v_{-})(1-v_{-})}{(1+v_{0}-v_{+}-v_{-})^2} \,.
\end{eqnarray}
The differential equation in $v_{0}$ turns out to be the simplest one,
namely
\begin{eqnarray}\label{diffeqv0}
\partial_{v_{0}} \tilde{\Phi}_{6}(v_{\pm},v_{0})
 =  \frac{v_{+}-v_{-}}{(v_{0} - v_{-})(v_{0} - v_{+})}
  \log \frac{(v_{0}-v_{+}v_{-})}{(1+v_{0}-v_{+}-v_{-}) v_{0}}
  \log \frac{(v_{0}-v_{+}v_{-}) v_{0} }{(1+v_{0}-v_{+}-v_{-}) v_{+} v_{-}} 
\,.
\end{eqnarray}
%
%{\bf Note sign flip in last equation.}
Using \eqns{xipmfromvi}{Jinvi}, 
%
\iffalse
% I don't think this identity is actually necessary!  -LD
and the identity
%
\begin{eqnarray}
\Li_{2}(1-x) + \Li_{2}\left(1-\frac{1}{x} \right) + \frac{1}{2} \log^2 (x)  = 0 \,,
\end{eqnarray}
\fi
%
it is easy to show that
\begin{eqnarray}
\partial_{v_{0}} L_{3}(x_{1+},x_{1-}) &=&
 \frac{1}{8} \frac{v_{+}-v_{-}}{(v_{0}-v_{-}) (v_{0}-v_{+})}
 \log^2 \left( \frac{ v_{+} v_{-}}{v_{0}^2} \right)\,,\\ 
\partial_{v_{0}} L_{3}(x_{3+},x_{3-}) &=& 
-\frac{1}{8} \frac{v_{+}-v_{-}}{(v_{0}-v_{-}) (v_{0}-v_{+})} 
\log^2 \left( \frac{ v_{+} v_{-} (1+v_{0}-v_{+}-v_{-})^2}
                   {(v_{0}-v_{+}v_{-})^2} \right)\,,\\ 
\partial_{v_{0}} L_{3}(x_{2+},x_{2-}) &=& 0 \,, \\
\partial_{v_{0}} J &=& 0 \,.
\end{eqnarray}
Hence $\tilde{\Phi}_{6}$ as defined in~\eqn{phitfullsolution}
satisfies~\eqn{diffeqv0}.
%{\bf Sign flips in last set of equations too.}
%
We have checked numerically that the differential equations with respect
to $v_{+}$ and $v_{-}$ are satisfied as well.

%%%%%%%%%%%%%%%%%%%%

\subsection{Symbol of $\tilde{\Phi}_{6}(u_{1},u_{2},u_{3})$}
\label{sect-symbol}

The notion of symbols has proven to be a useful tool for thinking
about transcendental functions appearing in $\cN=4$ SYM;
see ref.~\cite{Goncharov:2010jf} and references therein.

The symbol $[\tilde{\Phi}_{6}]$ of $\tilde{\Phi}_{6}$ is 
very simple, namely,
\begin{eqnarray}\label{phitsymbol}
[\tilde{\Phi}_{6}(u_{1},u_{2},u_{3})]
 = -[ \Omega^{(1)}(u_{1},u_{2},u_{3}) ] \otimes
\frac{x_{+} (1-x_{3 -})}{x_{-} (1-x_{3 +})}  + {\rm cyclic}\,,
\end{eqnarray}
where 
\begin{eqnarray}
[ \Omega^{(1)}(u_{1},u_{2},u_{3}) ]
= u_{1} \otimes u_{2} + u_{2} \otimes u_{1}
 - \sum_{i=1}^{3} u_{i} \otimes (1-u_{i})\,.
\end{eqnarray}
Note that the first of the three entries in $[\tilde{\Phi}_{6}]$ is
always either $u_1$, $u_2$ or $u_3$.  Because the $u_i$ are ratios
of the distances $x_{ij}^2$, using standard properties of the symbol
the first entry can always be expressed as a distance.  
This property has been argued to follow from the branch-cut
structure of loop integrals~\cite{Gaiotto:2011dt}. 

In order to see directly that~\eqn{phitsymbol} is the symbol of
\eqn{phitfullsolution} it is helpful to introduce to some projective variables $w_i \in \mathbb{CP}^1$ for $i=1,\ldots,6$.
Choosing homogeneous coordinates $w_i=(1,z_i)$, they coincide with the $z_i$ variables of \cite{Goncharov:2010jf}. We can represent the three cross-ratios as follows,
\begin{equation}
u_1 = \frac{(23)(56)}{(25)(36)}\,, \qquad u_2 = \frac{(34)(61)}{(36)(41)}\,, \qquad u_3 = \frac{(45)(12)}{(41)(52)}\,,
\end{equation}
where $(ij)=-(ji)=\epsilon_{ab} w_i^a w_j^b$. In terms of these variables $\Delta$ is a perfect square,
\begin{equation}
\Delta = \biggl[\frac{(12)(34)(56)+(23)(45)(61)}{(14)(25)(36)}\biggr]^2\,
\end{equation}
and all entries of the symbol factorize into two-brackets $(ij)$. Thus one can canonically represent the symbol as a sum of terms of the form
\begin{equation}
(ab)\otimes(cd)\otimes(ef)\,.
\end{equation}
Performing this on the symbol (\ref{phitsymbol}) and the symbol of (\ref{phitfullsolution}) one finds immediately the same expression.

One can easily check that the symbol 
of $\tilde{\Phi}_{6}$ is consistent
with the differential equation~(\ref{diffeqsinglevar1}) for
$\tilde{\Phi}_{6}$.  We simply replace the functions $\tilde{\Phi}_{6}$
and $\Omega^{(1)}$ in \eqn{diffeqsinglevar1} by their symbols,
and use the following simple identities,
\begin{eqnarray}
\partial_{u_{1}} \log  \frac{x_{+} (1-x_{1 -})}{x_{-} (1-x_{1 +})}
 &=& \frac{1-u_{1}-u_{2}-u_{3}}{u_{1} \sqrt{\Delta}}\,, \\
\partial_{u_{1}} \log  \frac{x_{+} (1-x_{2 -})}{x_{-} (1-x_{2 +})}
 &=& \frac{1-u_{1}-u_{2}+u_{3}}{(1-u_{1}) \sqrt{\Delta} }\,, \\
\partial_{u_{1}} \log  \frac{x_{+} (1-x_{3 -})}{x_{-} (1-x_{3 +})}
 &=& \frac{1-u_{1}+u_{2}-u_{3}}{(1-u_{1}) \sqrt{\Delta} }\,, 
\end{eqnarray} 
and the differentiation rule for symbols,
\begin{eqnarray}
\partial_{x} \, \left( a_{1} \otimes \ldots 
             \otimes a_{n-1} \otimes a_{n} \right)
 = \partial_{x} \log(a_{n}) \, \times \, a_{1} \otimes \ldots
             \otimes a_{n-1} \,.
\end{eqnarray}
This analysis can be used to justify the solution~(\ref{phitfullsolution}),
following ref.~\cite{Goncharov:2010jf}:
We have already seen that \eqn{phitfullsolution} has the correct symbol.
This leaves two ambiguities in $\Phi_{6}$, firstly where to place the 
branch cuts, and secondly the freedom to add constants multiplied by
functions of lower transcendentality than three.  The first ambiguity
is resolved by requiring that $\Phi_{6}$ be real-valued and smooth in
the entire Euclidean region $u_{i}>0$.  We have numerical evidence
that this is the case for $\Phi_{6}$ in~\eqn{phitfullsolution}.
The second ambiguity has to be fixed by other means.
The $\zeta_{2}$ term in \eqn{Om1} for $\Omega^{(1)}$,
which enters the differential equation~(\ref{link-phi-omega1}),
suggests the corresponding term in \eqn{phitfullsolution}.
We have also checked that the resulting formula is in agreement
with the parametric representation~(\ref{phi-feynman2}) for
several numerical values, which cover different regions according to
the signs of $\Delta$, $u_i-1$ and $u_1+u_2+u_3-1$.

%%%%%%%%%%%%%%%%%%%%%%%%%%%%%

\section{Conclusions and outlook}
\label{sect-conclusion}

In this paper, we have computed the six-dimensional one-loop on-shell 
scalar hexagon integral $\Phi_{6}$, giving its full kinematical
dependence in the Euclidean region. The result is a remarkably simple
formula, \eqn{phitfullsolution}.  Interestingly, its structure is
almost identical to that of the two-loop remainder function in
planar $\cN=4$ SYM~\cite{Goncharov:2010jf}, although the latter is
of transcendentality degree $4$, while $\Phi_{6}$ is of degree $3$.

Our calculation was based on the observation that $\Phi_{6}$ is
related to a known four-dimensional one-loop tensor hexagon integral
through first-order differential equations. The latter uniquely
determine the answer.  It is interesting to note that both the
two-loop remainder function and $\Phi_{6}$ are best expressed in terms
of a set of (redundant) variables $x_{i\pm}$. For $\Phi_{6}$, one is
led to these variables in a very natural way when solving the
aforementioned differential equations.  This approach should be very
helpful when computing other integrals of this kind.  In particular an
extension to degree five and six functions should provide valuable
insight into the structure of the remainder function at higher loops.
Another interesting extension of this work could be to consider the hexagon
integral with massive corners, which may give hints about
good sets of kinematic variables for amplitudes with
$n>6$ external legs.

The procedure for finding a relation between $\Omega^{(2)}$ and
$\Omega^{(1)}$ in ref.~\cite{Drummond:2010cz} was based on a Laplace
equation, which is second-order in nature, as are typical field
equations for bosonic fields.  On the other hand, fermionic field
equations are typically first order.  One might speculate that the
first-order relations~(\ref{intro-omega-phi-omega}) between
$\Omega^{(1)}, \Phi_{6}$ and $\Omega^{(2)}$ found in the present paper
could have an explanation based on supersymmetry.  What is somewhat
mysterious from this point of view is why the function $\Phi_6$ which
sits between $\Omega^{(1)}$ and $\Omega^{(2)}$ should have a full
cyclic symmetry, when neither $\Omega^{(1)}$ nor $\Omega^{(2)}$ do.

Finally, we comment that the fully off-shell version of $H$ has
a conventional conformal symmetry in addition to its dual conformal
symmetry.  This is the case simply because it is built from $\phi^3$
vertices, and $\phi^3$ theory in $D=6$ dimensions is classically
conformal.  By Fourier transforming the coordinate space conformal
generators $d, k^{\mu}$, and accounting for a change in conformal
dimension coming from the amputation of external legs, we find their
form in momentum space, acting on $H$,
\begin{eqnarray}
d  = \sum_{i=1}^{n} \Big\lbrack  p^{\nu}_{i} \partial_{i \nu} 
 + 2 \Big\rbrack \,, \quad
k^{\mu} &=& \sum_{i=1}^{n} 
\Big\lbrack -\frac{1}{2} p_{i}^{\mu} \partial_{i}^{\nu} \partial_{i \nu}
  + 2 \partial_{i}^{\mu}  + p_{i}^{\nu} \partial_{i \nu} \partial_{i}^{\mu} 
 \Big\rbrack  \,.
\end{eqnarray}
%
% D-dimensional version of this formula:
%\begin{eqnarray}
%d  = \sum_{i=1}^{n} \left[  p^{\nu}_{i} \partial_{i \nu}  + \frac{D}{2} -1 \right] \,, \quad
%k^{\mu} &=& \sum_{i=1}^{n} \left[ -\frac{1}{2} p_{i}^{\mu} \partial_{i}^{\nu} \partial_{i \nu}  + \left(\frac{D}{2} -1 \right) \partial_{i}^{\mu}  + p_{i}^{\nu} \partial_{i \nu} \partial_{i}^{\mu}  \right]  \,,
%\end{eqnarray}
%
Invariance under these operators then implies homogeneous second-order
differential equations.  If one takes some or all external legs on
shell, as in the case of $H$ (or $\Phi_{6}$), it can happen that the
action of the conformal generators becomes anomalous.

%%%%%%%%%%%%%

\section{Acknowledgments}
We thank M.~Spradlin and C.~Vergu for useful discussions.
This research was supported by the US Department of Energy under contract
DE--AC02--76SF00515.\\

{\bf Note added.} After this calculation was completed, we were informed by 
V.~Del Duca, C.~Duhr and V.~Smirnov of an independent computation of the
hexagon integral presented here, using a different method~\cite{DDS}.

\appendix

\section{A special case of $\Phi_{6}$}

The differential equations simplify considerably in the special
case $u_{3}=1$, for which $\sqrt{\Delta} = u_1-u_2$. (This is
true for $u_1 > u_2 $, which we can assume without loss of generality
since $\Phi_{6}$ is symmetric in $u_1 $ and $u_2 $.)
Starting from \eqn{diffeqsinglevar1}, and using
$\Omega^{(1)}(u_{2},1,u_{1}) = \Omega^{(1)}(1,u_{1},u_{2})$, 
we find
\begin{equation}
\partial_{u_{1}} \tilde{\Phi}_{6}(u_{1},u_{2},1)
= \frac{\Omega^{(1)}(u_{1},u_{2},1)}{1-u_{1}}
- \frac{\Omega^{(1)}(1,u_{1},u_{2})}{u_1(1-u_1)} \,.
\end{equation}
One can easily find the solution
\begin{equation}\label{solutiontwovar}
\Phi_{6}( u_1 , u_2 , 1 )
= \frac{\tilde\Phi_{6}( u_1 , u_2 , 1 )}{u_1-u_2}
= \frac{h(u_1,u_2) - h(u_2,u_1)}{u_1-u_2} \,,
% \big[ g(u_1) - g(u_2) + h(u_1,u_2) \big]\,,
\end{equation}
where
%
%\begin{eqnarray}
%g(u) &=& 2 \zeta_{2} \log u - \log (1 - u) \log^2 u - \log u  \, \PolyLog_{2}( 1 - u) -  2 \Log u \,  \PolyLog_{2} u +2 \, \PolyLog_{3} u  \,,\\
% JH: 2011-03-23: simplified expression for g(u) to
%g(u) &=&  \log u  \, (\zeta_{2} - \Li_{2} (u) )
%            + 2 \, \PolyLog_{3} (u)  \,,\\
%h(u,v) &=&  \log v  \, \PolyLog_{2} (1 - u ) - \Log u \, \PolyLog_{2} (1 - v) \% LD: 2011-03-23: simplified expression slightly further
%\end{eqnarray}
%
\begin{equation}
h(u_1,u_2) = \log u_1 \, ( \zeta_{2} - \Li_{2} (u_1) - \Li_2(1-u_2) )
               + 2 \, \Li_3(u_1) \,.
\label{def-h}
\end{equation}

%\subsection{MB representation}
%For numerical evaluation, or for taking limits analytically,
%the following MB representation can be useful,
%\begin{eqnarray}
%\Phi_{6}(u_1, u_2 , u_3 )  &=&  \int \frac{dz_{1,2,3}}{(2 \pi i)^3}\,  u_{1}^{z_4 } u_{2}^{-1 - z_1 - z_4} u_{3}^{z_1 - z_2 + z_4}
%  \Gamma(-z_1 ) \Gamma( -z_2 )  \Gamma^2(-z_4 ) \nonumber \\
%  && \times \Gamma^2(-z_1 + z_2 - z_4 ) \Gamma^2 (1 + z_1 + z_4 ) \Gamma(1 + z_1 - z_2 + 2 z_4 ) \,.
%\end{eqnarray}
%For example, one obtains
%\begin{equation}
%\Phi_{6}(x,x,x) = -  \log^3 x -  \frac{\pi^2}{2} \log x +O(x) \,.
%\end{equation}
%and
%\begin{eqnarray}
%\Phi_{6}(x, x , w ) &=& \log^2 x  \, \left[ - \frac{ \log w }{1-w} \right] +   \log x  \, \left[ -2 \,  \frac{ {\rm Li}_{2}( 1- w)  }{1-w} \right] \nonumber \\
%&& + \frac{1}{1-w} \left[ 
 %-2 \, {\rm Li}_{1,2}(1-w) + 2\, {\rm Li}_{3}(1- w )  - \log w \, {\rm Li}_{2}(1-w) 
%\right] + O(x) \,.
%\end{eqnarray}

%%%%%%%%%%%%%%

\section{Relations between $D=6$ integrals and $D=4$ tensor integrals}
\label{4d-6d-connection}

Here we give another example of a relation between a four-dimensional
tensor integral and a six-dimensional scalar integral. While the relation in
the main text involved a first-order differential operator, the relation we present
here is simply an equality of two integrals.

Let us consider the finite, dual conformal pentagon integral $\tilde{\Psi}$
\cite{ArkaniHamed:2010kv,Drummond:2010cz} that appears in the representation
of \cite{ArkaniHamed:2010kv} of one-loop MHV amplitudes in $\cN=4$ SYM.
Up to a normalization factor, it is given by
\begin{eqnarray}\label{eqPsitilde}
\tilde{\Psi} \propto  \int \frac{d^{4}x_{i}}{i \pi^2}
\frac{x_{ia}^2}{x_{2i}^2 x_{3i}^2 x_{5i}^2 x_{6i}^2 x_{8i}^2} \,,
\end{eqnarray}
where $x_{a}^{\mu}$ is defined as one of the two solutions to the four-cut 
conditions $x_{2a}^2 = x_{3a}^2 = x_{5a}^2 = x_{6a}^2 =0$.
As in the case of $\Omega^{(1)}$, the numerator factor makes the
integral IR finite. 

We remark that dual conformal transformations can be used to 
remove the $1/x_{8i}^2$ propagator, by letting $x^{\mu}_{8} \to \infty$, as in
ref. \cite{Broadhurst:1993ib}.
This is possible in this case because there are no light-like
constraints between $x^{\mu}_8$ and the neighboring $x^{\mu}_2$ and $x^{\mu}_6$.
In this way we obtain the equivalent 
integral 
\begin{eqnarray}
I = \int \frac{d^{4}x_{i}}{i \pi^2}
\frac{x_{ia}^2}{x_{2i}^2 x_{3i}^2 x_{5i}^2 x_{6i}^2} \,.
\end{eqnarray}
This integral is not dual conformally invariant, and is a function of $x_{25}^2, x_{26}^2, x_{35}^2, x_{36}^2$.
Up to a normalization factor, it equals the finite part of the
two-mass easy box integral~\cite{Bern:1993kr,Brandhuber:2004yw}
\begin{eqnarray}
I = \frac{-1}{x_{26}^2 + x_{35}^2 -x_{25}^2 - x_{36}^2}
 \left[ \Li_{2}(1 - \xi x_{26}^2 )+ \Li_{2}(1 - \xi x_{35}^2 )
      - \Li_{2}(1 - \xi x_{25}^2 )- \Li_{2}(1 - \xi x_{36}^2 ) \right]\,,
\end{eqnarray}
where $\xi = ( x_{26}^2 + x_{35}^2 -x_{25}^2 - x_{36}^2)
/(x_{26}^2 x_{35}^2-x_{25}^2 x_{36}^2)$.
Since the finite part of the one-loop MHV amplitude in $\mathcal{N}=4$ SYM is governed by this function
(the divergent parts correspond to one-mass and two-mass triangle integrals), 
this gives a very direct relation between six-dimensional integrals and four-dimensional amplitudes.

In order to see the relation of $I$ to a scalar integral in $D=6$
dimensions, one can introduce Feynman parameters, treating the
numerator $x_{ia}^2$ as an inverse propagator $1/
(x_{ia}^2)^{-1+\delta}$ with some auxiliary analytic regularization
$\delta$. Integrating out the Feynman parameter corresponding to this
inverse propagator and letting $\delta \to 0$, one readily obtains
\begin{eqnarray}
I = \int_{0}^{1} d\alpha_{2,3,5,6}
 \frac{\delta(1-\sum_{i=2,3,5,6} \alpha_i  )}
{\alpha_2 \alpha_5 x_{25}^2 +\alpha_3 \alpha_5 x_{35}^2
  + \alpha_2 \alpha_6 x_{26}^2 + \alpha_3 \alpha_6 x_{36}^2  }\,,
\end{eqnarray}
which is nothing else than the Feynman parametrization of the
following scalar integral in $D=6$ dimensions,
\begin{eqnarray}
I = \int \frac{d^{6}x_{i}}{i \pi^3} 
\frac{1}{x_{2i}^2 x_{3i}^2 x_{5i}^2 x_{6i}^2} \,.
\end{eqnarray}

We remark that by combining propagators pairwise 
(see appendix~\ref{appendix-wl}) and integrating out the
resulting finite bubble integral, one obtains a Wilson-loop 
type of representation for this
integral~\cite{Brandhuber:2007yx,Anastasiou:2011zk}.

%%%%%%%%%%%%%%%%%%%%%

\section{Wilson-loop representation of $\Phi_{6}$}
\label{appendix-wl}

%%%%%%%%%%%%%%%%%%%%%%%%%%%%%%%%%%%%%%%%%
\begin{figure}
\psfrag{x1}[cc][cc]{$\scriptstyle x_1$}
\psfrag{x2}[cc][cc]{$\scriptstyle x_2$}
\psfrag{x3}[cc][cc]{$\scriptstyle x_3$}
\psfrag{x4}[cc][cc]{$\scriptstyle x_4$}
\psfrag{x5}[cc][cc]{$\scriptstyle x_5$}
\psfrag{x6}[cc][cc]{$\scriptstyle x_6$}
\psfrag{xi1}[cc][cc]{$\scriptstyle y_1$}
\psfrag{xi3}[cc][cc]{$\scriptstyle y_3$}
\psfrag{xi5}[cc][cc]{$\scriptstyle y_5$}
\psfrag{equal}[cc][cc]{$=$}
\psfrag{two}[cc][cc]{$\scriptstyle 2$}
\psfrag{one}[cc][cc]{$\scriptstyle 1$}
\psfrag{a}[cc][cc]{(a)}
\psfrag{b}[cc][cc]{(b)}

 \centerline{
 {\epsfxsize12cm  \epsfbox{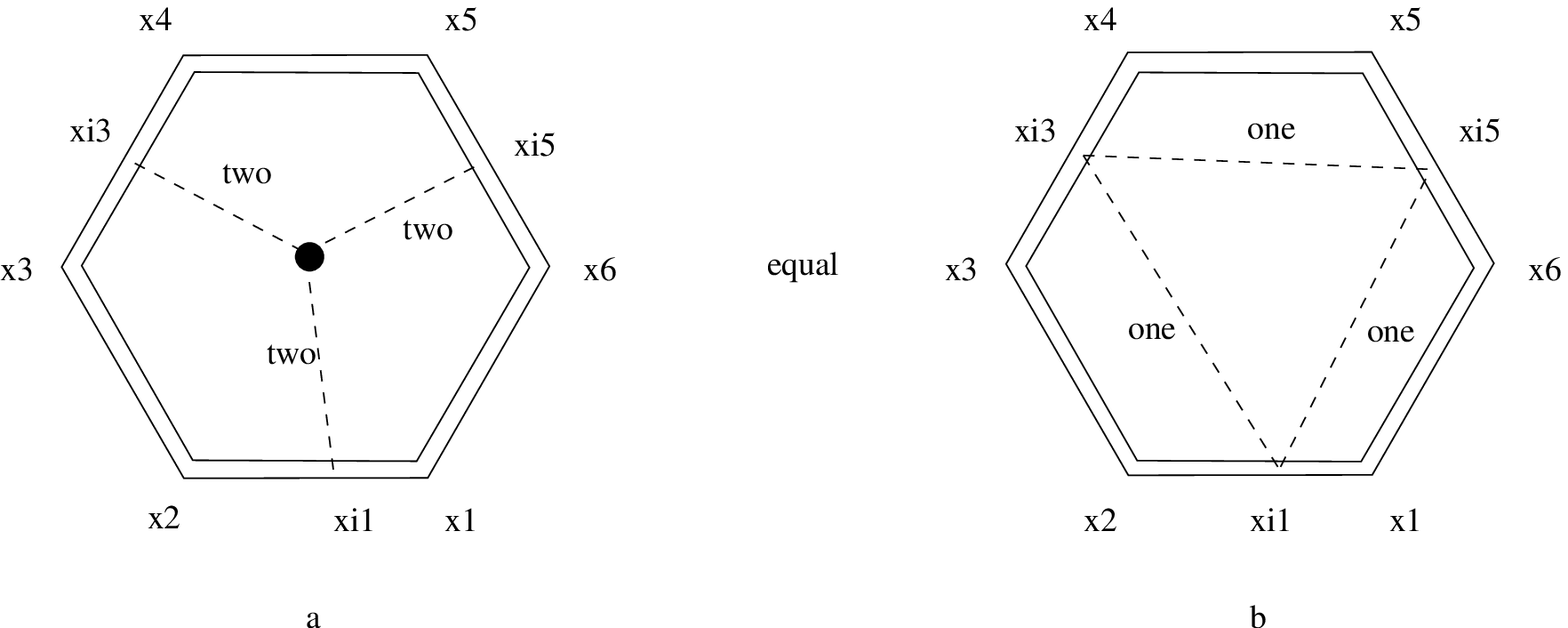}}
}
\caption{\small
Interpretation of the hexagon integral as a line integral,
according to \eqns{hex-wl-1}{hex-wl-2}.}
\label{fig-wl-representation}
\end{figure}
%%%%%%%%%%%%%%%%%%%%%%%%%%%%%%%%%%%%%%%%%

In section \ref{sect-preliminaries}, we explained how dual
conformal symmetry helps to obtain a convenient Feynman
parametrization for $H$, where in particular the number of
parameter integrals is equal to the degree of the function.
Here, we present a second way of exploiting dual conformal symmetry,
which in addition allows for an interpretation of $H$ as a Wilson-loop
integral.

Let us start from the definition of $H$ given in~\eqn{def-H}. It is
well-known that for on-shell integrals it is often desirable to
introduce Feynman parameters in steps, i.e.~to combine two adjacent
propagators at a time, using the formula
\begin{eqnarray}
\label{pairwiseprop}
\frac{1}{x_{1i}^2 x_{2i}^2 }
= \int_{0}^{1} d\xi_{1} \frac{1}{ [(y_{1} - x_{i})^2]^2 }\,,
\qquad x_{12}^2 =0\,,
\end{eqnarray}
where 
\begin{eqnarray}
\label{y1def}
y^{\mu}_{1}(\xi_{1}) = x^{\mu}_{1} (1-\xi_{1}) + x^{\mu}_{2} \xi_{1}\,.
\end{eqnarray}
For example, the two-mass easy box integral is ``easy'' precisely because
it contains two pairs of propagators separated by a massless leg;
\eqn{pairwiseprop} can be applied to each pair.

Repeating this procedure for the other two pairs of adjacent
propagators leads to
\begin{eqnarray}\label{hex-wl-1}
H = \int_{0}^{1} d\xi_{1,3,5}  \int \frac{d^{6}x_{i}}{i \pi^3}
\frac{1}{ [(y_{1} - x_{i})^2]^2 [(y_{3}- x_{i})^2]^2
          [(y_{5} - x_{i})^2]^2} \,,
\end{eqnarray}
where $y_3^\mu$ ($y_5^\mu$) is defined like $y_1^\mu$ in \eqn{y1def},
but with $i\rightarrow i+2$ ($i\rightarrow i+4$).
At the cost of having introduced three parameter integrals, the
innermost integral now depends on three ``effective propagators''
only, see Fig.~\ref{fig-wl-representation}(a). For a triangle integral,
however, dual conformal symmetry fixes the answer uniquely to be
a constant multiple of
$1/[(y_{1}-y_{3})^2 (y_{1}-y_{5})^2 (y_{3}-y_{5})^2]$.
The constant can be determined from a
boundary condition, e.g.~$y_{5} \to \infty$. This is nothing else
than the star-triangle (or uniqueness) relation
\cite{Kazakov:1984km}, of course.  Hence the answer is simply
\begin{eqnarray}\label{hex-wl-2}
H =  \int_{0}^{1} d\xi_{1,3,5}
 \frac{1}{ (y_{1}-y_{3})^2 (y_{1}-y_{5})^2 (y_{3}-y_{5})^2} \,,
\end{eqnarray}
which is depicted in see Fig.~\ref{fig-wl-representation}(b).
More explicitly, we have 
$(y_{1}-y_{3})^2 = x_{13}^2 \bar{\xi}_{1} \bar{\xi}_{3} 
 + x_{14}^2 {\xi}_{3} \bar{\xi}_{1} + x_{24}^2  {\xi}_{1}{\xi}_{3}$,
where $\bar{\xi} := 1-\xi$, etc.
In this form, the Feynman loop integral is reminiscent of a Wilson-loop
integral in the dual space of the $x_{i}$.
See ref.~\cite{Anastasiou:2011zk} for a similar discussion
of related integrals.

\bibliographystyle{nb}
\bibliography{hex6d}

\begin{thebibliography}{10}
\ifx\href\asklfhas\newcommand{\href}[2]{#2}\fi
\ifx\arxivref\asklfhas\newcommand{\arxivref}[2]{\href{http://arxiv.org/abs/#1}%
{#2}}\fi
\ifx\doiref\asklfhas\newcommand{\doiref}[2]{\href{http://dx.doi.org/#1}{#2}}\fi
\raggedright
\small
\parskip 0pt

%%CITATION = HEP-TH/0607160;%%
\bibitem{Drummond:2006rz}
J.~M.~Drummond, J.~Henn, V.~A.~Smirnov and E.~Sokatchev,
\textit{``{Magic identities for conformal four-point integrals}''},
\textsf{\doiref{10.1088/1126-6708/2007/01/064}{JHEP~0701,~064~(2007)}},
\texttt{\arxivref{hep-th/0607160}{hep-th/0607160}}.

%%CITATION = HEP-TH 0610248;%%
\bibitem{Bern:2006ew}
Z.~Bern, M.~Czakon, L.~J.~Dixon, D.~A.~Kosower and V.~A.~Smirnov,
\textit{``The four-loop planar amplitude and cusp anomalous dimension in
  maximally supersymmetric Yang-Mills theory''},
\textsf{\doiref{10.1103/PhysRevD.75.085010}{Phys.~Rev.~D75,~085010~(2007)}},
\texttt{\arxivref{hep-th/0610248}{hep-th/0610248}}.

%%CITATION = 0709.2368;%%
\bibitem{Drummond:2007cf}
J.~M.~Drummond, J.~Henn, G.~P.~Korchemsky and E.~Sokatchev,
\textit{``{On planar gluon amplitudes/Wilson loops duality}''},
\textsf{\doiref{10.1016/j.nuclphysb.2007.11.007}{Nucl.~Phys.~B795,~52~(2008)}},
\texttt{\arxivref{0709.2368}{arxiv:0709.2368}}.

%%CITATION = 0705.0303;%%
\bibitem{Alday:2007hr}
L.~F.~Alday and J.~M.~Maldacena,
\textit{``{Gluon scattering amplitudes at strong coupling}''},
\textsf{\doiref{10.1088/1126-6708/2007/06/064}{JHEP~0706,~064~(2007)}},
\texttt{\arxivref{0705.0303}{arxiv:0705.0303}}.

%%CITATION = 0707.0243;%%
\bibitem{Drummond:2007aua}
J.~M.~Drummond, G.~P.~Korchemsky and E.~Sokatchev,
\textit{``{Conformal properties of four-gluon planar amplitudes and Wilson
  loops}''},
\textsf{\doiref{10.1016/j.nuclphysb.2007.11.041}{Nucl.~Phys.~B795,~385~(2008)}%
},
\texttt{\arxivref{0707.0243}{arxiv:0707.0243}}.

%%CITATION = 0707.1153;%%
\bibitem{Brandhuber:2007yx}
A.~Brandhuber, P.~Heslop and G.~Travaglini,
\textit{``{MHV Amplitudes in {$\mathcal{N}=\mathord{}$4} Super Yang--Mills and
  Wilson Loops}''},
\textsf{\doiref{10.1016/j.nuclphysb.2007.11.002}{Nucl.~Phys.~B794,~231~(2008)}%
},
\texttt{\arxivref{0707.1153}{arxiv:0707.1153}}.

%%CITATION = 1009.2225;%%
\bibitem{Mason:2010yk}
L.~Mason and D.~Skinner,
\textit{``{The Complete Planar S-matrix of {$\mathcal{N}=\mathord{}$4} SYM as a
  Wilson Loop in Twistor Space}''},
\textsf{\doiref{10.1007/JHEP12(2010)018}{JHEP~1012,~018~(2010)}},
\texttt{\arxivref{1009.2225}{arxiv:1009.2225}}.

%%CITATION = 1010.1167;%%
\bibitem{CaronHuot:2010ek}
S.~Caron-Huot,
\textit{``{Notes on the scattering amplitude / Wilson loop duality}''},
\texttt{\arxivref{1010.1167}{arxiv:1010.1167}}.

\bibitem{Belitsky:2011zm}
A.~Belitsky, G.~Korchemsky and E.~Sokatchev,
\textit{``{Are scattering amplitudes dual to super Wilson loops?}''},
\texttt{\arxivref{1103.3008}{arxiv:1103.3008}}.

\bibitem{Gaiotto:2011dt}
D.~Gaiotto, J.~Maldacena, A.~Sever and P.~Vieira,
\textit{``{Pulling the straps of polygons}''},
\texttt{\arxivref{1102.0062}{arxiv:1102.0062}}.

%%CITATION = 0712.1223;%%
\bibitem{Drummond:2007au}
J.~M.~Drummond, J.~Henn, G.~P.~Korchemsky and E.~Sokatchev,
\textit{``{Conformal Ward identities for Wilson loops and a test of the duality
  with gluon amplitudes}''},
\textsf{\doiref{10.1016/j.nuclphysb.2009.10.013}{Nucl.~Phys.~B826,~337~(2010)}%
},
\texttt{\arxivref{0712.1223}{arxiv:0712.1223}}.

%%CITATION = HEP-TH/0309040;%%
\bibitem{Anastasiou:2003kj}
C.~Anastasiou, Z.~Bern, L.~J.~Dixon and D.~A.~Kosower,
\textit{``{Planar amplitudes in maximally supersymmetric Yang-Mills theory}''},
\textsf{\doiref{10.1103/PhysRevLett.91.251602}{Phys.~Rev.~Lett.~91,~251602~(20%
03)}},
\texttt{\arxivref{hep-th/0309040}{hep-th/0309040}}.

%%CITATION = HEP-TH 0505205;%%
\bibitem{Bern:2005iz}
Z.~Bern, L.~J.~Dixon and V.~A.~Smirnov,
\textit{``Iteration of planar amplitudes in maximally supersymmetric Yang-Mills
  theory at three loops and beyond''},
\textsf{\doiref{10.1103/PhysRevD.72.085001}{Phys.~Rev.~D72,~085001~(2005)}},
\texttt{\arxivref{hep-th/0505205}{hep-th/0505205}}.

\bibitem{Cachazo:2006tj}
F.~Cachazo, M.~Spradlin and A.~Volovich,
\textit{``{Iterative structure within the five-particle two-loop amplitude}''},
\textsf{\doiref{10.1103/PhysRevD.74.045020}{Phys.Rev.~D74,~045020~(2006)}},
\texttt{\arxivref{hep-th/0602228}{hep-th/0602228}}.

%%CITATION = HEP-TH/0604074;%%
\bibitem{Bern:2006vw}
Z.~Bern, M.~Czakon, D.~A.~Kosower, R.~Roiban and V.~A.~Smirnov,
\textit{``{Two-loop iteration of five-point {$\mathcal{N}=\mathord{}$4}
  super-Yang-Mills amplitudes}''},
\textsf{\doiref{10.1103/PhysRevLett.97.181601}{Phys.~Rev.~Lett.~97,~181601~(20%
06)}},
\texttt{\arxivref{hep-th/0604074}{hep-th/0604074}}.

%%CITATION = 0803.1465;%%
\bibitem{Bern:2008ap}
Z.~Bern, L.~J.~Dixon, D.~A.~Kosower, R.~Roiban, M.~Spradlin, C.~Vergu and
  A.~Volovich,
\textit{``{The Two-Loop Six-Gluon MHV Amplitude in Maximally Supersymmetric
  Yang-Mills Theory}''},
\textsf{\doiref{10.1103/PhysRevD.78.045007}{Phys.~Rev.~D78,~045007~(2008)}},
\texttt{\arxivref{0803.1465}{arxiv:0803.1465}}.

%%CITATION = 0805.4832;%%
\bibitem{Cachazo:2008hp}
F.~Cachazo, M.~Spradlin and A.~Volovich,
\textit{``{Leading Singularities of the Two-Loop Six-Particle MHV
  Amplitude}''},
\textsf{\doiref{10.1103/PhysRevD.78.105022}{Phys.~Rev.~D78,~105022~(2008)}},
\texttt{\arxivref{0805.4832}{arxiv:0805.4832}}.

%%CITATION = 0808.1054;%%
\bibitem{Spradlin:2008uu}
M.~Spradlin, A.~Volovich and C.~Wen,
\textit{``{Three-Loop Leading Singularities and BDS Ansatz for Five
  Particles}''},
\textsf{\doiref{10.1103/PhysRevD.78.085025}{Phys.~Rev.~D78,~085025~(2008)}},
\texttt{\arxivref{0808.1054}{arxiv:0808.1054}}.

%%CITATION = 1004.5381;%%
\bibitem{Henn:2010ir}
J.~M.~Henn, S.~G.~Naculich, H.~J.~Schnitzer and M.~Spradlin,
\textit{``{More loops and legs in Higgs-regulated {$\mathcal{N}=\mathord{}$4}
  SYM amplitudes}''},
\textsf{\doiref{10.1007/JHEP08(2010)002}{JHEP~1008,~002~(2010)}},
\texttt{\arxivref{1004.5381}{arxiv:1004.5381}}.

%%CITATION = 0803.1466;%%
\bibitem{Drummond:2008aq}
J.~M.~Drummond, J.~Henn, G.~P.~Korchemsky and E.~Sokatchev,
\textit{``{Hexagon Wilson loop = six-gluon MHV amplitude}''},
\textsf{\doiref{10.1016/j.nuclphysb.2009.02.015}{Nucl.~Phys.~B815,~142~(2009)}%
},
\texttt{\arxivref{0803.1466}{arxiv:0803.1466}}.

%%CITATION = 0705.1864;%%
\bibitem{Bern:2007ct}
Z.~Bern, J.~J.~M.~Carrasco, H.~Johansson and D.~A.~Kosower,
\textit{``{Maximally supersymmetric planar Yang--Mills amplitudes at five
  loops}''},
\textsf{\doiref{10.1103/PhysRevD.76.125020}{Phys.~Rev.~D76,~125020~(2007)}},
\texttt{\arxivref{0705.1864}{arxiv:0705.1864}}.

%%CITATION = 1005.2902;%%
\bibitem{Henn:2010kb}
J.~M.~Henn,
\textit{``{Scattering amplitudes on the Coulomb branch of
  {$\mathcal{N}=\mathord{}$4} super Yang-Mills}''},
\textsf{\doiref{10.1016/j.nuclphysbps.2010.08.042}{Nucl.~Phys.~Proc.~Suppl.~20%
5-206,~193~(2010)}},
\texttt{\arxivref{1005.2902}{arxiv:1005.2902}}.

\bibitem{Dennen:2010dh}
T.~Dennen and Y.-t.~Huang,
\textit{``{Dual Conformal Properties of Six-Dimensional Maximal Super
  Yang-Mills Amplitudes}''},
\textsf{\doiref{10.1007/JHEP01(2011)140}{JHEP~1101,~140~(2011)}},
\texttt{\arxivref{1010.5874}{arxiv:1010.5874}}.

\bibitem{CaronHuot:2010rj}
S.~Caron-Huot and D.~O'Connell,
\textit{``{Spinor Helicity and Dual Conformal Symmetry in Ten Dimensions}''},
\texttt{\arxivref{1010.5487}{arxiv:1010.5487}}.

\bibitem{Henn:2011xk}
J.~M.~Henn,
\textit{``{Dual conformal symmetry at loop level: massive regularization}''},
\texttt{\arxivref{1103.1016}{arxiv:1103.1016}}.

%%CITATION = 0911.5332;%%
\bibitem{DelDuca:2009au}
V.~Del~Duca, C.~Duhr and V.~A.~Smirnov,
\textit{``{An Analytic Result for the Two-Loop Hexagon Wilson Loop in
  {$\mathcal{N}=\mathord{}$4} SYM}''},
\textsf{\doiref{10.1007/JHEP03(2010)099}{JHEP~1003,~099~(2010)}},
\texttt{\arxivref{0911.5332}{arxiv:0911.5332}}.

%%CITATION = 1003.1702;%%
\bibitem{DelDuca:2010zg}
V.~Del~Duca, C.~Duhr and V.~A.~Smirnov,
\textit{``{The Two-Loop Hexagon Wilson Loop in {$\mathcal{N}=\mathord{}$4}
  SYM}''},
\textsf{\doiref{10.1007/JHEP05(2010)084}{JHEP~1005,~084~(2010)}},
\texttt{\arxivref{1003.1702}{arxiv:1003.1702}}.

%%CITATION = 1006.5703;%%
\bibitem{Goncharov:2010jf}
A.~B.~Goncharov, M.~Spradlin, C.~Vergu and A.~Volovich,
\textit{``{Classical Polylogarithms for Amplitudes and Wilson Loops}''},
\textsf{\doiref{10.1103/PhysRevLett.105.151605}{Phys.~Rev.~Lett.~105,~151605~(%
2010)}},
\texttt{\arxivref{1006.5703}{arxiv:1006.5703}}.

%%CITATION = 0802.2065;%%
\bibitem{Bartels:2008ce}
J.~Bartels, L.~N.~Lipatov and A.~Sabio~Vera,
\textit{``{BFKL Pomeron, Reggeized gluons and Bern-Dixon-Smirnov
  amplitudes}''},
\textsf{\doiref{10.1103/PhysRevD.80.045002}{Phys.~Rev.~D80,~045002~(2009)}},
\texttt{\arxivref{0802.2065}{arxiv:0802.2065}}.

\bibitem{Bartels:2008sc}
J.~Bartels, L.~Lipatov and A.~Sabio~Vera,
\textit{``{{$\mathcal{N}=\mathord{}$4} supersymmetric Yang Mills scattering
  amplitudes at high energies: The Regge cut contribution}''},
\textsf{\doiref{10.1140/epjc/s10052-009-1218-5}{Eur.~Phys.~J.~C65,~587~(2010)}%
},
\texttt{\arxivref{0807.0894}{arxiv:0807.0894}}.

%%CITATION = 1008.2965;%%
\bibitem{Drummond:2010mb}
J.~M.~Drummond and J.~M.~Henn,
\textit{``{Simple loop integrals and amplitudes in {$\mathcal{N}=\mathord{}$4}
  SYM}''},
\texttt{\arxivref{1008.2965}{arxiv:1008.2965}}.

\bibitem{Drummond:2010cz}
J.~M.~Drummond, J.~M.~Henn and J.~Trnka,
\textit{``{New differential equations for on-shell loop integrals}''},
\texttt{\arxivref{1010.3679}{arxiv:1010.3679}}.

\bibitem{ArkaniHamed:2010kv}
N.~Arkani-Hamed, J.~L.~Bourjaily, F.~Cachazo, S.~Caron-Huot and J.~Trnka,
\textit{``{The All-Loop Integrand For Scattering Amplitudes in Planar
  {$\mathcal{N}=\mathord{}$4} SYM}''},
\textsf{\doiref{10.1007/JHEP01(2011)041}{JHEP~1101,~041~(2011)}},
\texttt{\arxivref{1008.2958}{arxiv:1008.2958}}.

%%CITATION = HEP-TH/9611127;%%
\bibitem{Bern:1996ja}
Z.~Bern, L.~J.~Dixon, D.~C.~Dunbar and D.~A.~Kosower,
\textit{``{One-loop self-dual and {$\mathcal{N}=\mathord{}$4}
  super-Yang-Mills}''},
\textsf{\doiref{10.1016/S0370-2693(96)01676-0}{Phys.~Lett.~B394,~105~(1997)}},
\texttt{\arxivref{hep-th/9611127}{hep-th/9611127}}.

\bibitem{Bern:1993kr}
Z.~Bern, L.~J.~Dixon and D.~A.~Kosower,
\textit{``{Dimensionally regulated pentagon integrals}''},
\textsf{\doiref{10.1016/0550-3213(94)90398-0}{Nucl.Phys.~B412,~751~(1994)}},
\texttt{\arxivref{hep-ph/9306240}{hep-ph/9306240}}.

%%CITATION = PHLTA,B307,132;%%
\bibitem{Broadhurst:1993ib}
D.~J.~Broadhurst,
\textit{``{Summation of an infinite series of ladder diagrams}''},
\textsf{\doiref{10.1016/0370-2693(93)90202-S}{Phys.~Lett.~B307,~132~(1993)}}.

\bibitem{smirnov2006feynman}
V.~Smirnov,
\textit{``{Feynman Integral Calculus}''},
Springer Verlag (2006).

\bibitem{ArkaniHamed:2010gh}
N.~Arkani-Hamed, J.~L.~Bourjaily, F.~Cachazo and J.~Trnka,
\textit{``{Local Integrals for Planar Scattering Amplitudes}''},
\texttt{\arxivref{1012.6032}{arxiv:1012.6032}}.

\bibitem{DDS}
V.~Del~Duca, C.~Duhr and V.~A.~Smirnov,
\texttt{\arxivref{to~appear}{to~appear}}.

%%CITATION = HEP-TH/0407214;%%
\bibitem{Brandhuber:2004yw}
A.~Brandhuber, B.~J.~Spence and G.~Travaglini,
\textit{``{One-loop gauge theory amplitudes in {$\mathcal{N}=\mathord{}$4}
  super Yang-Mills from MHV vertices}''},
\textsf{\doiref{10.1016/j.nuclphysb.2004.11.023}{Nucl.~Phys.~B706,~150~(2005)}%
},
\texttt{\arxivref{hep-th/0407214}{hep-th/0407214}}.

\bibitem{Anastasiou:2011zk}
C.~Anastasiou and A.~Banfi,
\textit{``{Loop lessons from Wilson loops in {$\mathcal{N}=\mathord{}$4}
  supersymmetric Yang-Mills theory}''},
\textsf{\doiref{10.1007/JHEP02(2011)064}{JHEP~1102,~064~(2011)}},
\texttt{\arxivref{1101.4118}{arxiv:1101.4118}}.

\bibitem{Kazakov:1984km}
D.~Kazakov,
\textit{``{The method of uniqueness, a new powerful technique for multiloop
  calculations}''},
\textsf{\doiref{10.1016/0370-2693(83)90816-X}{Phys.Lett.~B133,~406~(1983)}}.

\end{thebibliography}

\end{document}